\documentclass[aps,prb,preprint,groupedaddress]{revtex4}
\usepackage{graphicx}
\usepackage{bm}
\usepackage{color}
\usepackage{enumerate}
\usepackage{amsmath}
\usepackage{dcolumn}
\begin{document}
%
\title{%
Insulator-to-Superconductor Transition upon Electron Doping in a BiS$_{2}$-Based Superconductor Sr$_{1-x}$La$_{x}$FBiS$_{2}$
}
\author{H. Sakai$^{1}$}
\author{D. Kotajima$^{1}$}
\author{K. Saito$^{1}$}
\author{H. Wadati$^{2}$}
\author{Y. Wakisaka$^{3}$}
\author{M. Mizumaki$^{4}$}
\author{K. Nitta$^{4}$}
\author{Y. Tokura$^{1,5}$}
\author{S. Ishiwata$^{1}$}
\affiliation{%
$^{1}$Department of Applied Physics, University of Tokyo, Tokyo 113-8656, Japan\\
$^{2}$Department of Applied Physics and Quantum-Phase Electronics Center (QPEC), University of Tokyo, Tokyo 113-8656, Japan\\
$^{3}$Photon Factory, Institute of Materials Structure Science, High Energy Accelerator Research Organization, Tsukuba 305-0801, Japan\\
$^{4}$Japan Synchrotron Radiation Research Institute, SPring-8, Hyogo 679-5198, Japan\\
$^{5}$RIKEN Center for Emergent Matter Science (CEMS), Wako 351-0198, Japan
}
%
%
\begin{abstract}
Effects of electron doping on the BiS$_{2}$-based superconductors Sr$_{1-x}$La$_{x}$FBiS$_{2}$ ($0\!\le\!x\!\le\!0.6$) have been investigated using the systematically synthesized polycrystals by means of x-ray diffraction, x-ray absorption spectroscopy, transport and thermodynamic measurements.
The pristine compound is a band insulator with the BiS$_{2}$ layer, which accommodates electron carriers through the La substitution for Sr, as evidenced by the change in x-ray absorption spectra reflecting the occupancy of Bi 6$p$ orbitals.
With increasing the carrier density, the resistivity progressively decreases and a bad metallic state appears for $x\!\ge\!0.45$, where bulk superconductivity manifests itself below approximately 3 K.
The value of $T_{\rm c}$ gradually increases with decreasing $x$ from 0.6 to 0.45 and immediately decreases down to zero at the critical concentration of $x\!\sim\!0.4$, resulting in an insulator-superconductor transition highly sensitive to the carrier density.
Thermodynamic measurements furthermore have revealed the possible enhancement of the superconducting coupling strength as the insulating phase is approached.
The obtained superconducting phase diagram is markedly different from the broad dome-shaped superconducting phase previously reported for $R$O$_{1-x}$F$_{x}$BiS$_{2}$ ($R$: rare-earth ion), suggesting a strong influence of the blocking layer on the superconductivity.
Instead all these features are similar to those observed in Li-intercalated ZrNCl superconductor, except for the critical electron concentration of as low as 6\% in the latter compound.
For the present superconductor, notably, the existence of hole-type carriers has been indicated in the normal state from the Hall effect measurements.
The Sr$_{1-x}$La$_{x}$FBiS$_{2}$ system providing the phase diagram for the rigid-band doping in the BiS$_{2}$ layer would be another prototypical example of superconductivity derived from a doped layered band insulator, hosting both hole-like and electron-like Fermi surfaces.
\end{abstract}
%
\maketitle
%
\section{Introduction}
%
Exotic superconductivity with high-$T_{\rm c}$ has been often discovered in doped layered compounds, such as cuprates\cite{Bednorz1986}, $\beta$-$T$NCl ($T$=Zr, Hf)\cite{Yamanaka1998Nature}, and iron pnictides\cite{Kamihara2008JACS}.
Recently, materials with BiS$_{2}$ layers have attracted much attention as a new family of layered superconductors.
Superconductivity was first reported in Bi$_{4}$S$_{4}$O$_{3}$\cite{Mizuguchi2012PRB} ($T_{\rm c}\!\sim\!4.5$ K), followed by $R$O$_{1-x}$F$_{x}$BiS$_{2}$ ($R$=La, Ce, Pr, Nd and Yb)\cite{Mizuguchi2012JPSJ, Xing2012PRB, Yazici2012PhilosMag}, where the maximum $T_{\rm c}$ reaches $\sim$10 K around $x$=0.5 for $R$=La.
A characteristic crystal structure common to these materials is the BiS$_{2}$ layers consisting of the Bi and S ions positioned alternately on the square lattice, which are separated by insulating blocking layers.
The superconductivity has been believed to reside in the BiS$_{2}$ layer, when electron-type carriers are doped.
For the $R$O$_{1-x}$F$_{x}$BiS$_{2}$ compounds, for instance, the partial chemical substitution of O ions with F ions in the blocking layer allows the tuning of electron concentration in the BiS$_{2}$ layer and hence the critical temperature.
Such a stacking structure with mixed anions is reminiscent of so-called 1111-type Fe-based superconductors.\cite{Ishida2009JPSJ}
The variety of the blocking layer has been rapidly increasing, which includes Sr$_{0.5}$La$_{0.5}$FBiS$_{2}$\cite{Lin2013PRB}, La$_{1-x}M_{x}$OBiS$_{2}$ ($M$=Ti, Zr, Hf, and Th)\cite{Yazici2013PRB}, and Bi$_{3}$O$_{2}$S$_{3}$\cite{Phelan2013JACS}.
%
\par
%
Theoretically, the undoped BiS$_{2}$-based material was predicted to be a band insulator.
First-principles calculation suggested that the conduction band primarily consists of Bi 6$p_{x}$ and 6$p_{y}$ orbitals arising from the BiS$_{2}$ layer, hybridized with S 3$p$ orbitals.\cite{Usui2012PRB}
Assuming electron doping within a rigid-band scheme, the two-orbital model was suggested as a minimal model that considers only the Bi 6$p$ orbitals as the relevant bands crossing the Fermi surface.\cite{Usui2012PRB, Martins2013PRB}
It was predicted that the topology of the Fermi surface varies depending on the carrier concentration.
In particular, hole pockets appear around the $\Gamma$ (0,0) and $M$ ($\pi$,$\pi$) points, when the doping level is increased up to $\sim$0.5.
There, the Fermi surfaces exhibit nesting and the associated spin fluctuation was proposed as a candidate of the superconducting pairing.\cite{Martins2013PRB}
On the other hand, some theoretical works indicated the electron-phonon coupling as another pairing mechanism\cite{Li2013EPL}, which is enhanced in the vicinity of the ferroelectric and/or charge-density-wave transitions\cite{Yildirim2013PRB}.
However, this has not been supported by experiments yet.\cite{Lee2013PRB}
%
\par
%
The superconducting phase diagram was experimentally revealed for $R$O$_{1-x}$F$_{x}$BiS$_{2}$ ($R$=La, Ce, and Nd).\cite{Deguchi2013EPL, Xing2012PRB, Demura2013JPSJ}
Except for the difference in $T_{\rm c}$, the overall features of variation in $T_{\rm c}$ are similar for all the compounds; a broad dome-shaped superconducting phase insensitive to the carrier density manifests itself around $x$=0.5.
However, the behavior of $T_{\rm c}$ around the phase boundary to a non-superconducting phase has been still ambiguous.
The carrier concentration above which the value of $T_{\rm c}$ is non-zero appears to significantly differ depending on the materials; it is approximately $x$=0.1 for $R$=La, whereas $x$=0.25 for $R$=Ce.
There have been also several complicated transport properties in addition to the emergence of the superconductivity.
For LaO$_{1-x}$F$_{x}$BiS$_{2}$, while the resistivity slightly decreases with increasing $x$ up to 0.2, it starts to steeply increase with further increasing $x$ and exhibits semiconducting behavior above $x$=0.5.\cite{Deguchi2013EPL}
For CeO$_{1-x}$F$_{x}$BiS$_{2}$, on the other hand, the pristine compound is a bad metal and the resistivity monotonically increases with increasing $x$.\cite{Xing2012PRB}
%
\par
%
Such a puzzling behavior may be attributed to the change in band structure via the partial substitution of F ions for O ions.
It was recently suggested that the band gap is substantially enlarged by fluorine doping, since the top of the valence band mainly originates from O 2$p$ orbitals, which tends to sink as fluorine is doped.\cite{Suzuki2013Procedia}
To reveal the intrinsic dependence of $T_{\rm c}$ on the carrier density, it is essential to afford the carriers to the BiS$_{2}$ layer without modifying the original band structure.
For this purpose, we have here focused on the recently-synthesized SrFBiS$_{2}$ as a pristine compound.\cite{Lei2013IC}
Importantly, a first-principles calculation predicted the top of the valence band consists mainly of the S 3$p$ orbitals, where the density of states of the Sr and F orbitals in the blocking layer is negligibly small.
This would provide an ideal arena to realize the rigid-band carrier doping into the BiS$_{2}$ layer by substituting Sr ions with La ones.
Very recently, the superconductivity was actually discovered for Sr$_{0.5}$La$_{0.5}$FBiS$_{2}$\cite{Lin2013PRB}, which has put a strong emphasis upon the importance of the extensive study about the dependence on carrier concentration.
%
\par
%
In this study, we aim to elucidate the genuine phase diagram associated with the superconductivity residing in the BiS$_{2}$ layer.
We have systematically synthesized polycrystalline Sr$_{1-x}$La$_{x}$FBiS$_{2}$ ($0\!\le\!x\!\le\!0.6$) samples and revealed the detailed electronic properties of both the superconducting and normal states by performing the x-ray diffraction, x-ray absorption, transport and thermodynamic measurements.
The obtained phase diagram is characterized by an insulator-superconductor transition with a steep phase boundary at the critical concentration around $x$=0.4.
This is markedly different from that for $R$O$_{1-x}$F$_{x}$BiS$_{2}$, indicating the strong impact of the blocking layer on the superconductivity.
Unusual enhancement of the superconducting coupling strength as well as $T_{\rm c}$ has been suggested as the carrier concentration is reduced toward the critical point.
We discuss these superconducting properties in comparison with the Li-intercalated ZrNCl compound.
%
\section{Experimental}
%
Sr$_{1-x}$La$_{x}$FBiS$_{2}$ polycrystals ($0.25\!\le\! x\!\le\!0.6$) were synthesized by the following steps of solid state reaction and high-pressure annealing.
First, the stoichiometric mixtures of SrF$_{2}$, SrS, LaF$_{3}$, LaS$_{3}$, Bi and S powders were pressed into pellets and heated in an evacuated quartz tube at 600$^{\circ}$C for 10 h.
For $x$=0, the starting materials were the stoichiometric SrF$_{2}$, SrS, and the prereacted Bi$_{2}$S$_{3}$ powders.
The pellets for $x\!>\!0$ were ground in an Ar-filled globe box and sealed in a gold capsule.
A small amount of titanium powder was added at the both ends of the sample to avoid oxidization.
Then, the capsule was treated at $\sim$2.0 GPa and $\sim$600$^{\circ}$C for 1h, using a conventional cubic anvil-type high-pressure apparatus.
%
\par 
%
The obtained compounds were characterized at room temperature by 2$\theta$-$\theta$ powder x-ray diffraction (XRD) using Cu-K$\alpha$ radiation.
The x-ray absorption spectroscopy measurements at the Bi $L_{1}$ edge were carried out at room temperature by synchrotron radiation at the beamline BL01B1 of SPring-8, Japan.
The magnetization was measured with a superconducting quantum interference device (Quantum Design).
The four-probe resistivity ($\rho$), Hall coefficients ($R_{\rm H}$), and specific heat ($C$) were measured using Physical Property Measurement System (Quantum Design).
For the $C$ measurements at 0 T (superconducting state) and 9 T (normal state), the thermometers on the calorimeter puck were calibrated at both fields.
The addenda signal was determined before mounting the sample.
%
\section{Results and Discussion}
%
\subsection{Variation in lattice structure and Bi valence with La doping}
\begin{figure}
\begin{center}
\includegraphics[width=\linewidth]{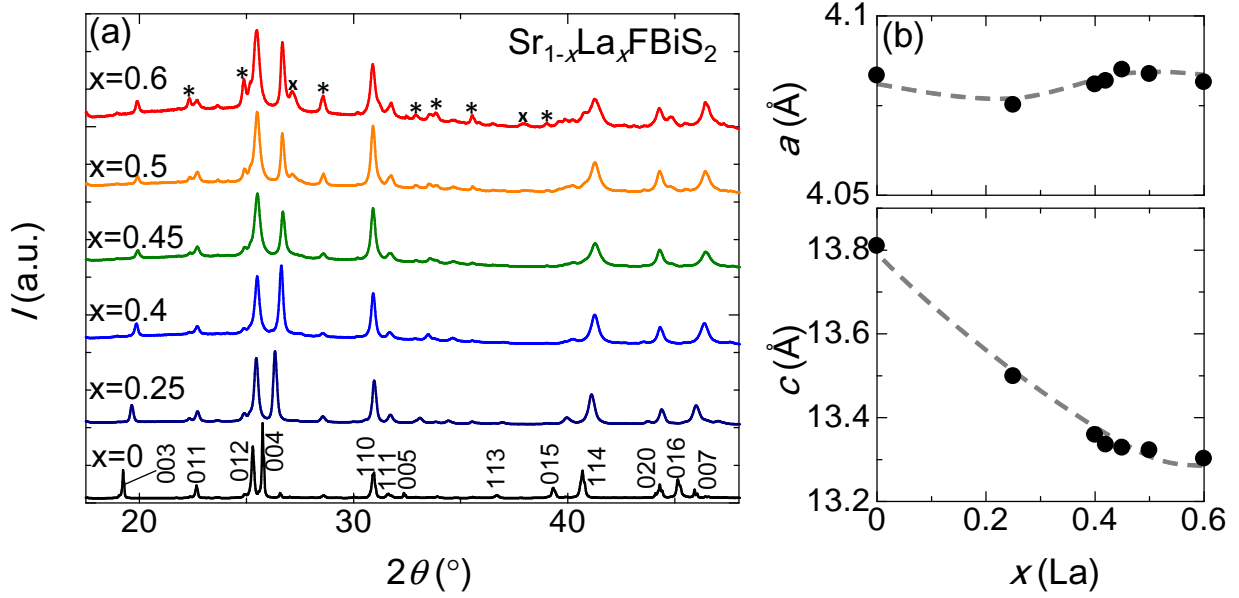}
\caption{\label{fig:XRD}(Color online) (a) X-ray diffraction profile at room temperature for Sr$_{1-x}$La$_{x}$FBiS$_{2}$ (0$\le\! x\!\le$0.6). $\ast$ and $\times$ denote the impurity peaks of Bi$_{2}$S$_{3}$ and Bi metal, respectively. The numbers are Miller indices based on the space group $P4/nmm$. (b) Lattice constants $a$ and $c$ as a function of $x$ (La concentration). The dashed lines are guides to eyes.}
\end{center}
\end{figure}

%
Figure \ref{fig:XRD}(a) displays the XRD profiles for Sr$_{1-x}$La$_{x}$FBiS$_{2}$ ($0\!\le\! x\!\le\!0.6$) at room temperature.
All the main peaks are assigned to Bragg reflections calculated for the tetragonal space group $P4/nmm$, which is consistent with the previous reports.\cite{Lei2013IC, Lin2013PRB}
While minute impurity peaks arising from Bi$_{2}$S$_{3}$ are discernible for all $x$, those from Bi metal impurity show up for $x\!\ge\!0.5$.
Note that the amount of both impurity phases largely increases for $x$=0.6, indicating the La solubility limit may be almost reached.
In Fig. \ref{fig:XRD}(b), we show the $x$ dependence of lattice constants $a$ (in-plane) and $c$ (out-of-plane) estimated by Le Bail analysis.
The $a$ value is almost unchanged as a function of $x$, whereas the $c$ value systematically decreases with increasing $x$, which indicates the formation of solid solution.
For $x\!\ge\!0.5$, however, the reduction in $c$ with an increase in $x$ starts to saturate. 
This again implies the solubility limit close to $x\!\sim\!0.6$ in the present compounds.
%
%
\par
\begin{figure}
\begin{center}
\includegraphics[width=.8\linewidth]{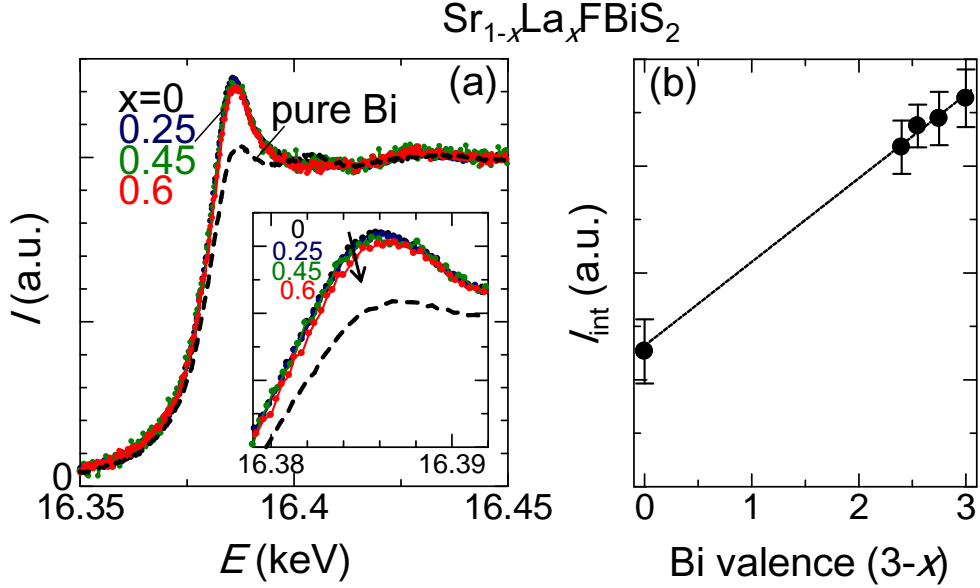}
\caption{\label{fig:XAS}(Color online) (a) Bi $L_{1}$ ($2s\!\rightarrow\!6p$) edge x-ray absorption spectra for Sr$_{1-x}$La$_{x}$FBiS$_{2}$ ($x$=0, 0.25, 0.45 and 0.6) and pure Bi. The inset shows the enlarged view around the maximum of the peak structure. (b) The integrated intensities of the $L_{1}$ edge spectra as a function of nominal Bi valence ($3\!-\!x$). The data for Bi metal is plotted as the zero-valence state. The error bars denote the uncertainty in integration range.}
\end{center}
\end{figure}

 
%
To clarify the variation in valence state of Bi ion with La substitution, we have carried out the x-ray absorption spectroscopy (XAS).
Figure \ref{fig:XAS}(a) shows the XAS spectra for $x$=0, 0.25, 0.45 and 0.6 around the Bi $L_{1}$ ($2s\!\rightarrow\!6p$) edge, which should be sensitive to the change in the 6$p$ states caused by electron doping into Bi$^{3+}$ ions.
The data for pure Bi are also displayed as a reference corresponding to the zero-valence state.
Each Bi $L_{1}$ XAS spectrum consists of a clear single peak.
As shown in the inset of Fig. \ref{fig:XAS}(a), the intensity of the peak gradually decreases with increasing $x$, reflecting the reduction in empty $6p$ states.
For quantitative comparison, we plotted in Fig. \ref{fig:XAS}(b) the integrated intensity of the Bi $L_{1}$ edge for each $x$ versus the nominal Bi valence, i.e., $3\!-\!x$.
We here assume that La$^{3+}$ substitution for Sr$^{2+}$ affords one electron to the BiS$_{2}$ layer.
The integrated intensity deceases almost linearly with increasing (decreasing) $x$ (nominal Bi valence) and the linear extrapolation coincides with the data for pure Bi.
This can be a firm evidence that electrons are effectively transferred into the BiS$_{2}$ conducting layer by La substitution in the SrF blocking layer.
%
%
\subsection{Electronic phase diagram for the Sr$_{1-x}$La$_{x}$FBiS$_{2}$ system}
\begin{figure}
\begin{center}
\includegraphics[width=.7\linewidth]{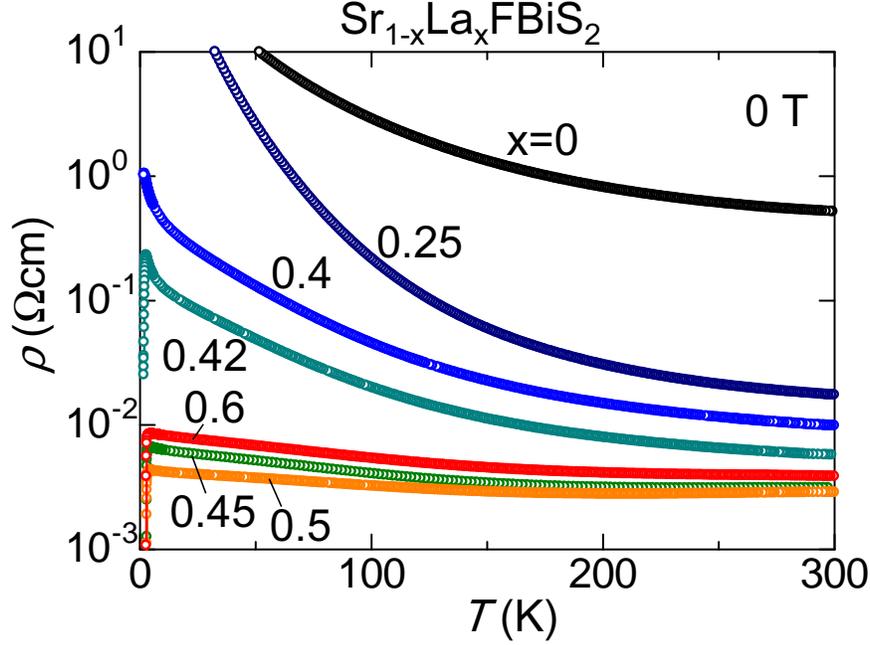}
\caption{\label{fig:RT}(Color online) Temperature dependence of resistivity for Sr$_{1-x}$La$_{x}$FBiS$_{2}$ (0$\le\! x\!\le$0.6) at 0 T.}
\end{center}
\end{figure}

\begin{figure}
\begin{center}
\includegraphics[width=.7\linewidth]{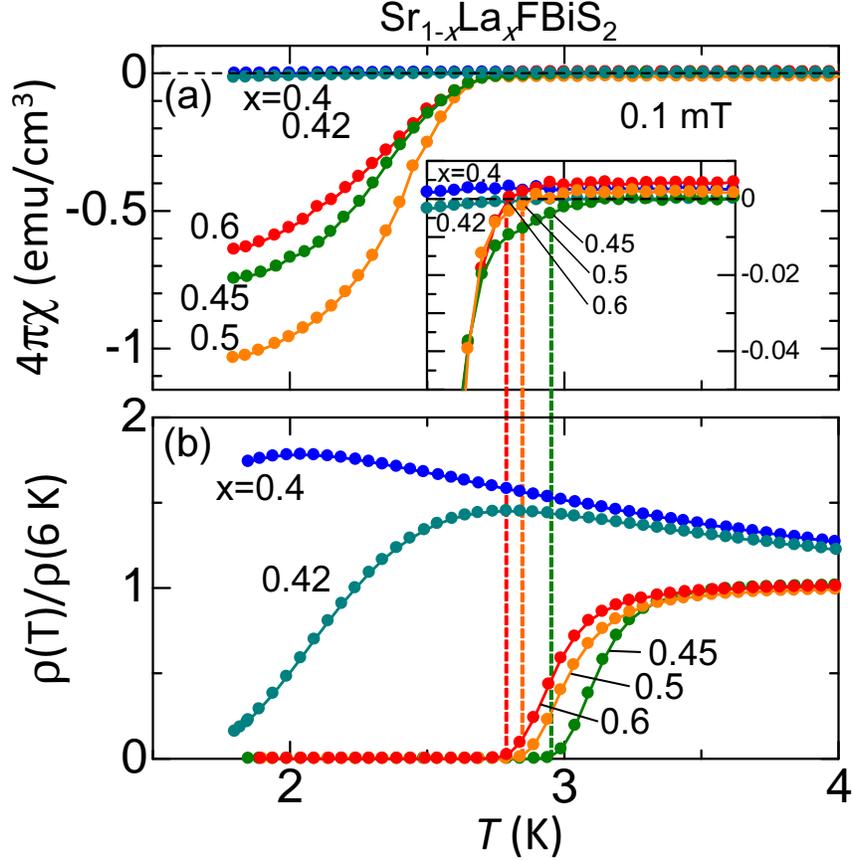}
\caption{\label{fig:RTMT}(Color online) Temperature dependence of (a) dc magnetic susceptibility (at 0.1 mT after zero field cooling) and resistivity (at 0 T)  for Sr$_{1-x}$La$_{x}$FBiS$_{2}$ (0.4$\le\! x\!\le$0.6) below 4 K. The resistivity is normalized by the value at 6 K for each compound. Inset in (a) is the enlarged view close to $T_{\rm c}$. The vertical dashed lines correspond to the zero-resistivity temperature for each superconducting compound.}
\end{center}
\end{figure}

\begin{figure}
\begin{center}
\includegraphics[width=.65\linewidth]{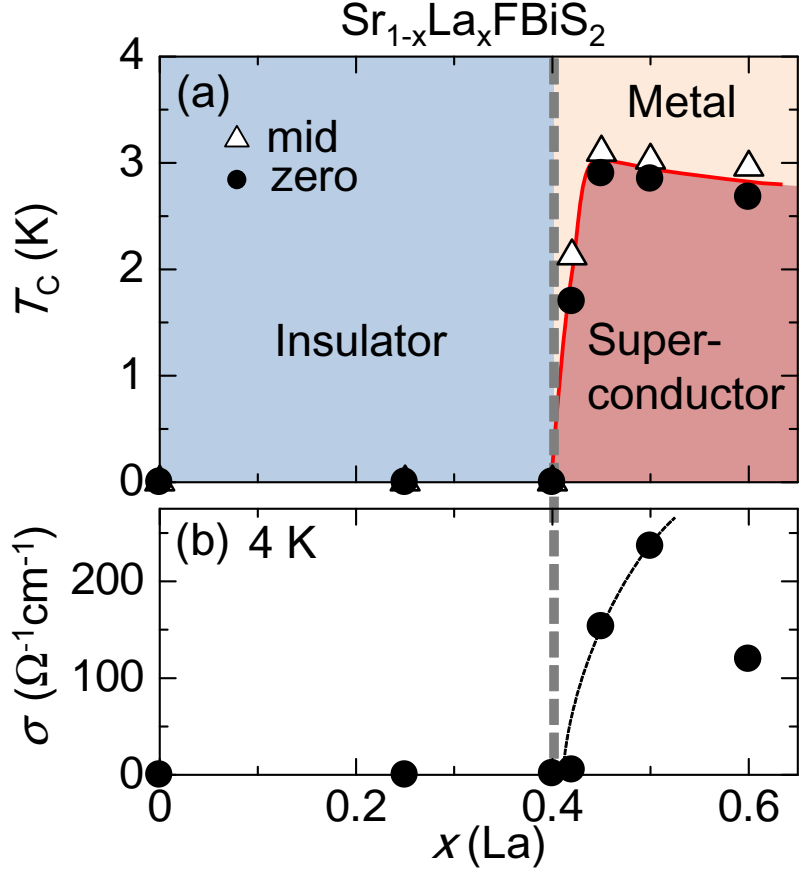}
\caption{\label{fig:phase}(Color online) (a) Electronic phase diagram and variation of $T_{\rm c}$ as a function of $x$ (La concentration). Filled circles (zero) and open triangle (mid) denote $T_{\rm c}$ determined by the temperature where the resistivity is 10\% and 50\% of that for the normal state, respectively. The solid curve is a guide to the eyes. (b) The normal-state conductivity at 4 K ($> T_{\rm c}$) as a function of $x$. The dashed curve is a guide to the eyes.}
\end{center}
\end{figure}

%
Figure \ref{fig:RT} presents the temperature profile of resistivity at 0 T for Sr$_{1-x}$La$_{x}$FBiS$_{2}$ ($0\!\le\! x\!\le\!0.6$).
The undoped ($x$=0) compound shows insulating behavior in the entire temperature region.
Above 150 K, the resistivity exhibits the thermal-activation type temperature dependence with the activation energy of $\sim$25 meV.
With increasing La concentration ($x$), the resistivity systematically decreases.
For $x$=0.4 and 0.42, the resistivity value at 300 K is significantly reduced (by nearly two orders of magnitude), but it still increases with decreasing temperature.
At low temperatures below 25 K, in particular, the conductivity obeys the $T^{0.5}$ dependence, which signals the weak Anderson localization effect.\cite{Lee1985RMP}
The sample with $x$=0.45 exhibits metallic behavior above 200 K, while it robustly shows a weak upturn at low temperatures probably due to the localization effect.
For $x$=0.5 and 0.6, the overall feature of resistivity is similar to that for $x$=0.45, although the change in absolute value is not systematic; the resistivity value for $x$=0.6 is slightly larger than those for $x$=0.45 and 0.5, which may be owing to the larger volume fraction of impurity phases for $x$=0.6 [Fig. \ref{fig:XRD}(a)].
A clear superconducting transition is discernible at around 3 K for $x\!\ge\!0.42$ (See Fig. \ref{fig:RTMT} for details).
Note here that this value of $x$ nicely corresponds to the insulator-metal transition of the normal state.
In fact, the conductivity of the normal state at zero temperature is non-zero for $x\!\ge\!0.45$ (and nearly zero for $x$=0.42), when the temperature profile above $T_{\rm c}$ is fitted to the $T^{0.5}$ dependence [See also Fig. \ref{fig:phase}(b)].
Thus, an insulator-to-metal transition takes place upon electron doping in the BiS$_{2}$ layer and the superconducting transition occurs in the doping-induced metallic phase at low temperatures.
%
%
\par 
%
To reveal the details of the superconducting state for the present system, we show in Figs. \ref{fig:RTMT}(a) and (b) the temperature dependence of dc magnetic susceptibility and resistivity below 4 K, respectively.
The magnetization was measured at 0.1 mT after zero-field cooling.
All the superconducting samples except for $x$=0.42 exhibit substantial diamagnetic responses at the lowest temperature.
The apparent shielding fractions for $x\!\ge\!0.45$ exceed 60\% at 1.8 K without demagnetizing-field correction, which ensures that the superconductivity is a bulk property.
As shown in the inset to Fig. \ref{fig:RTMT}(a), the temperature where the diamagnetic signal starts to steeply evolve well coincides with the zero-resistivity temperature for each $x$ (denoted as a dashed vertical line).
The diamagnetic signal for $x$=0.42 would be significant below 1.8 K, since the resistivity is still on the way of decreasing to zero at 1.8 K.
%
\par 
%
The $x$ dependence of $T_{\rm c}$ is noticeable in the resistivity data shown in Fig. \ref{fig:RTMT}(b).
The highest $T_{\rm c}$ was observed for $x$=0.45, where the zero-resistivity temperature is $\sim$3.0 K.
With decreasing $x$ from 0.45, $T_{\rm c}$ rapidly drops and a clear superconducting transition disappears for $x$=0.4.
When $x$ is increased from $x$=0.45, on the other hand, the superconducting state robustly remains up to $x$=0.6 with the zero-resistivity temperature gradually decreasing down to $\sim$2.8 K.
We here note that the resistivity for $x\!\ge\!0.45$ shows barely metallic behavior above $T_{\rm c}$.
Below $\sim$6 K, in fact, the temperature profile of resistivity deviates from the $T^{-0.5}$ dependence characteristic of weak localization, which might be caused by the superconducting fluctuation and/or disorder effects inherent in polycrystalline specimens.
%
%
\par 
%
In Fig. \ref{fig:phase}(a), we summarize the electronic phase diagram for the Sr$_{1-x}$La$_{x}$FBiS$_{2}$ system as functions of $x$ and temperature.
The values of $T_{\rm c}$ corresponding to the zero-resistivity and the mid-point of the transition are defined as the temperatures where the resistivity is 10\% and 50\% of the normal-state resistivity, respectively.
The superconducting phase manifests itself above $x$=0.4, where $T_{\rm c}$ shows unusual $x$ dependence.
As $x$ is reduced from 0.6 to 0.45, the zero-resistivity $T_{\rm c}$ gradually increases from 2.8 K to 3.0 K.
With further decreasing $x$ toward the phase boundary to the insulator, $T_{\rm c}$ steeply drops and the superconducting transition disappears at $x$=0.4.
The resultant phase diagram is characterized by an insulator-superconductor transition with a sharp phase boundary at approximately $x$=0.4, which significantly differs from the broad superconducting dome observed for $R$O$_{1-x}$F$_{x}$BiS$_{2}$ ($R$=La, Ce, and Nd).\cite{Deguchi2013EPL, Demura2013JPSJ, Xing2012PRB}
The critical concentration for the present compound ($x\!\sim\!0.4$) is also much higher than those for $R$O$_{1-x}$F$_{x}$BiS$_{2}$ ($x\!\sim\!0.1$-0.25).
The difference in blocking layer thus strongly influences the emergence of the superconducting phase.
As mentioned in the introductory part, this may be caused by the modification of the band structure upon the chemical substitution in the blocking layer.
Since the electronic states arising from the SrF blocks are situated away from the Fermi energy\cite{Lei2013IC}, the present result is regarded as intrinsic phase diagram of the electron-doped BiS$_{2}$-layer compound with a rigid band.
%
\par 
%
Another important feature is the doping-induced insulator-metal transition in the normal state, as mentioned above (Fig. \ref{fig:RT}).
To make it clear, in Fig. \ref{fig:phase}(b), we show the conductivity for the normal state at 4 K versus $x$.
For $x\!\le\!0.4$, the conductivity is virtually zero, whereas it is non-zero and continuously increases with increasing $x$ above $x$=0.4.
A sudden decrease in conductivity for $x$=0.6 may reflect the enhanced volume fraction of the impurity phases.
In the normal state, thus, the insulating phase is replaced above $x$=0.4 by the metallic one that condenses into the superconducting state at low temperatures.
It is worth noting that similar phase diagram was reported for the Li$_{x}$ZrNCl system.\cite{Taguchi2006PRL}
This compound shows an insulator-metal transition upon Li intercalation, although the critical carrier concentration is much lower ($\sim$0.06) than the present system.
In the superconducting phase, furthermore, $T_{\rm c}$ increases with reducing carrier density on the verge of insulator-superconductor transition, similarly to the present system.
As an origin of such an increase in $T_{\rm c}$, it has recently been proposed that the pairing interaction strength is enhanced with approaching a band insulator.\cite{Kasahara2009PRL}
Interestingly, such a tendency is also likely in the Sr$_{1-x}$La$_{x}$FBiS$_{2}$ compounds, as has been indicated by the thermodynamic measurements ($vide$ $infra$).
%
%
\subsection{Thermodynamic properties}
\begin{figure}
\begin{center}
\includegraphics[width=.7\linewidth]{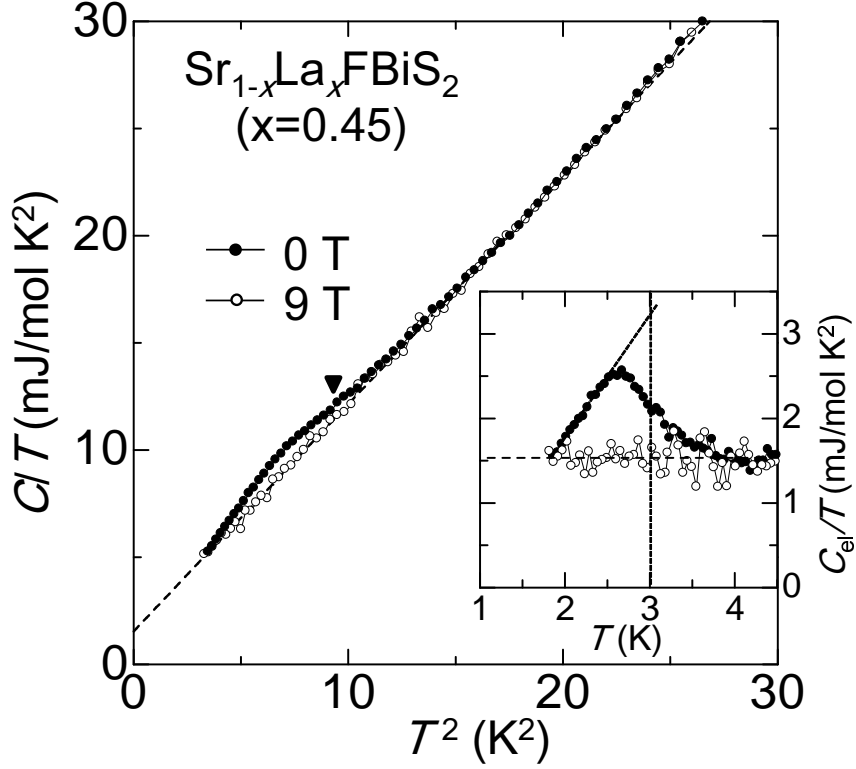}
\caption{\label{fig:CTx045}Temperature dependence of specific heat $C$ at 0 T (superconducting state) and 9 T (normal state) for Sr$_{1-x}$La$_{x}$FBiS$_{2}$ ($x$=0.45), plotted in the form of $C/T$ versus $T^{2}$. The closed triangle denotes the anomaly arising from the superconducting transition. The dashed line is the linear fit ($C/T=\gamma_{\rm n}+\beta T^{2}$) to the low-temperature data at 9 T. Inset shows the electronic part of the specific heat at 0 T and 9 T, obtained after subtracting the phononic contribution determined by the data at 9 T ($C_{\rm el}/T=C/T-\beta T^{2}$).}
\end{center}
\end{figure}

%
To further study the superconducting properties for the Sr$_{1-x}$La$_{x}$FBiS$_{2}$ system, we have systematically measured the specific heat.
Figure \ref{fig:CTx045} displays the temperature dependence of total specific heat $C$ for $x$=0.45, plotted as $C/T$ versus $T^{2}$.
The data at 0 T and 9 T correspond to the superconducting and normal states, respectively.
An anomaly relevant to the superconducting transition is discernible at $T^{2}$=9 K$^{2}$ in the zero-field curve, reflecting the bulk superconductivity in this compound.
The data for the normal state, on the other hand, are well fitted to the conventional relation: $C/T\!=\!\gamma_{\rm n}\!+\!\beta T^{2}$, as shown by a dashed line.
The best-fitted result gives the normal-state specific-heat coefficient of $\gamma_{\rm n}\!=\!1.6$ mJ/mol K$^{2}$, which is roughly in accordance with the value previously reported for Sr$_{0.5}$La$_{0.5}$FBiS$_{2}$.\cite{Lin2013PRB}
After subtracting the phononic contribution ($\beta T^{2}$), we have presented the electronic specific heat $C_{\rm el}$ in the inset of Fig. \ref{fig:CTx045}.
At 0 T, a clear jump manifests itself at the superconducting transition.
Taking the entropy balance into consideration, we have obtained $T_{\rm c}\!=\!3.0$ K and the specific-heat jump $\Delta C/\gamma_{\rm n}T_{\rm c}\!=\!1.1$.
Although the size of the observed specific-heat jump is less than that of the BCS weak-coupling limit (=1.43), it can be comparable to the BCS value when considering the apparent shielding fraction of $\sim$75\% [Fig. \ref{fig:RTMT}(a)], .
%
%
\par
\begin{figure}
\begin{center}
\includegraphics[width=.65\linewidth]{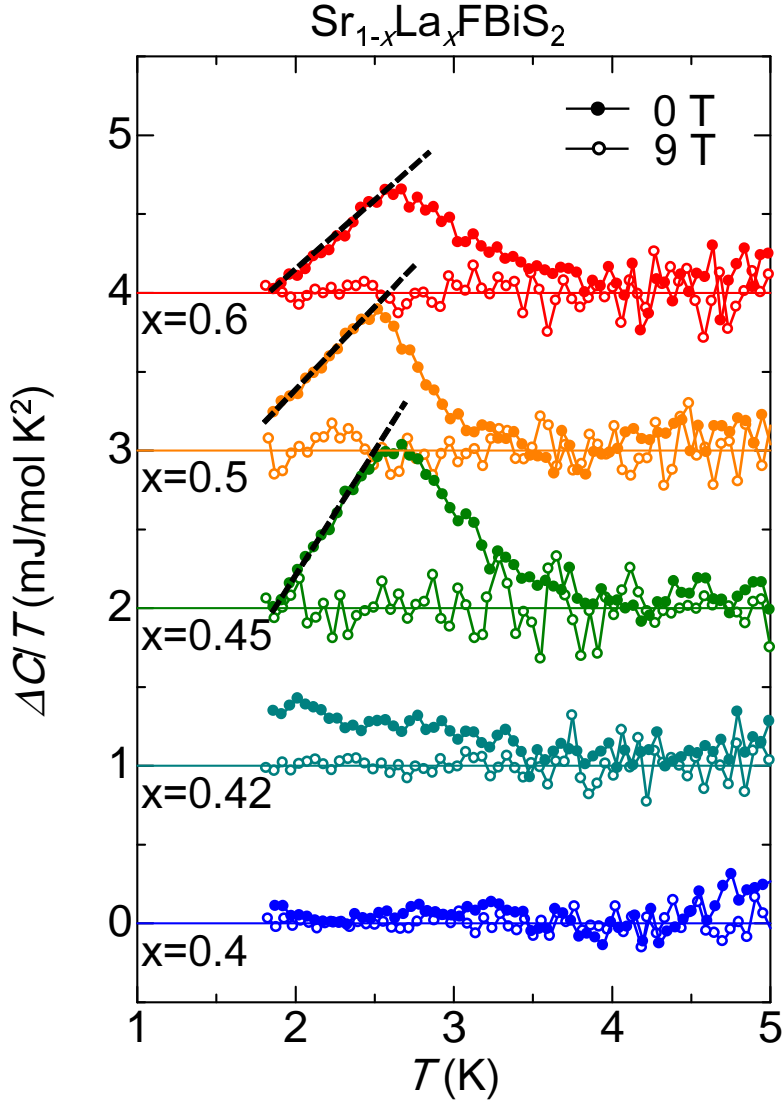}
\caption{\label{fig:CTall}(Color online) Temperature profiles of $\Delta C/T=[C(H)-C({\rm 9 T})]/T$ (the difference in electronic specific heat from the normal-state value) for $0.4\!\le\! x \!\le\! 0.6$ at $H$=0 and 9 T. The curves are shifted for clarity as denoted by horizontal solid lines.}
\end{center}
\end{figure}

 
%
Figure \ref{fig:CTall} displays the temperature profiles of $\Delta C/T\!=\![C(H,T)\!-\!C(H\!=\!9\ {\rm T},T)]/T$ at $H$=0 and 9 T for $0.4\!\le\!x\!\le\!0.6$.
The value of $\Delta C/T$ at 0 T is the difference in electronic specific heat between the superconducting and normal states.
We here subtracted the 9 T data by fitting them to a polynomial.
For $x$=0.4 and 0.42, no distinct anomaly in specific heat was observed in the measured temperature range.
A weak upturn observed for $x$=0.42 near the lowest temperature is reminiscent of its relatively broad superconducting transition.
For $x\!\ge\!0.45$, on the other hand, a clear jump is discernible at the superconducting transition.
Noteworthy is that the temperature profile of $\Delta C/T$ at 0 T varies depending on $x$; the gradient of $\Delta C/T$ below $T_{\rm c}$ gradually decreases with increasing $x$, as highlighted by the dashed lines.
In fact, the gradient of $\Delta C/T$ for $x$=0.5 decreases by approximately 35\% compared to $x$=0.45, in spite of the similar broadness of the transition.
The gradient seems to be further reduced for $x$=0.6, although the superconducting transition is broader than those for $x$=0.45 and 0.5.
Concomitantly, the size of specific-heat jump at $T_{\rm c}$ also gradually decreases with increasing $x$.
These facts suggest the change in either superconducting coupling strength or electronic specific heat with $x$, i.e., the carrier concentration, when considering the empirical theory, the so-called $\alpha$ model \cite{Padamsee1973}, that has a parameter of coupling strength $\alpha\!=\!\Delta_{0}/k_{\rm B}T_{\rm c}$, where $\Delta_{0}$ is the superconducting gap size at 0 K.
We here roughly estimate the values of $\gamma_{\rm s}\!=\!\gamma_{\rm n}\!-\!\gamma_{0}$, where $\gamma_{0}$ is the residual electronic specific-heat coefficient from the non-superconducting part of the sample.
Based on the values of $\gamma_{\rm n}$ and apparent volume fraction, we have obtained $\gamma_{\rm s}$=1.2 and 1.3 mJ/mol K$^{2}$ for $x$=0.45 and 0.5, respectively, indicating the difference in electronic specific heat may be less than 10\% between these two compounds.\cite{gamma}
In this case, the observed increase in gradient of $\Delta C/T$ below $T_{\rm c}$ with decreasing $x$ would essentially reflect the increase in coupling strength $\alpha$.
In the present system, the superconducting coupling strength is thus likely to be promoted toward the phase boundary to the insulator, in parallel with $T_{\rm c}$.
%
\par 
%
As mentioned above, the similar reinforcement of the pairing interaction was first suggested for Li$_{x}$ZrNCl.\cite{Kasahara2009PRL}
There, the possible pairing mediated by the spin fluctuation was proposed even in the doped band insulators, since the spin susceptibility as well as the pairing interaction was found to be enhanced with approaching a band insulator phase.
In the Sr$_{1-x}$La$_{x}$FBiS$_{2}$ compound, the pairing mechanism is still unclear.
The systematic measurements of the phonon spectra would be quite important to elucidate whether the conventional electron-phonon coupling alone can explain the observed $T_{\rm c}$ variation, or an exotic paring mechanism related to electron correlation is applicable to the present case as well.
%
%
\subsection{Hall effects}
\begin{figure}
\begin{center}
\includegraphics[width=.65\linewidth]{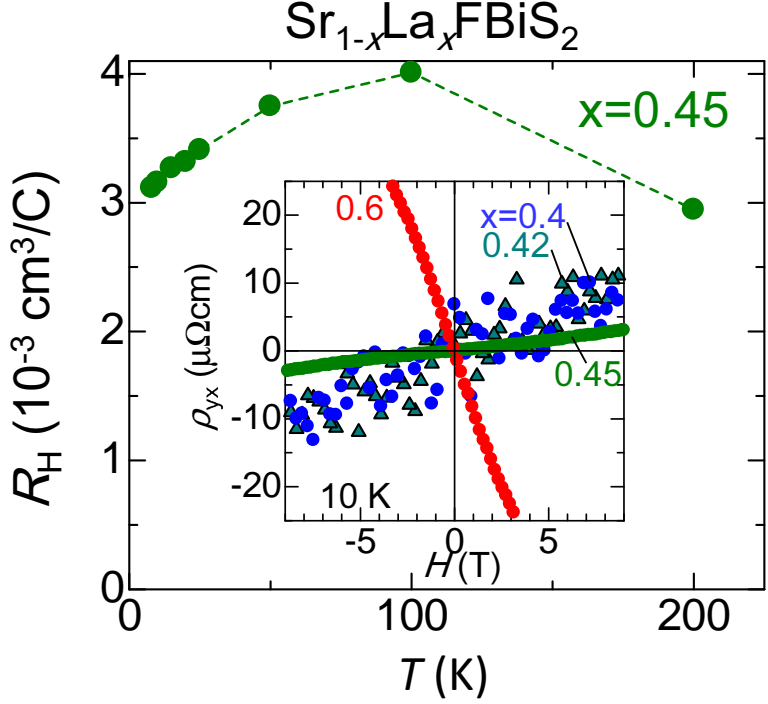}
\caption{\label{fig:Hall}(Color online) Temperature profile of Hall coefficient $R_{\rm H}$ for $x$=0.45. The inset shows the Hall resistivity $\rho_{yx}$ versus field at 10 K for $0.4\!\le\! x \!\le\! 0.6$.}
\end{center}
\end{figure}

%
To study the normal state in detail, we have carried out the Hall effect measurements on the samples that show a superconducting transition at low temperatures.
Figure \ref{fig:Hall} presents the temperature dependence of the Hall coefficient ($R_{\rm H}$) for $x$=0.45.
$R_{\rm H}$ was estimated by fitting linearly the Hall resistivity ($\rho_{yx}$) data up to $|H|\!\le\!9$ T, although the $H$ profile of $\rho_{yx}$ is slightly curved as shown in the inset of Fig. \ref{fig:Hall}.
Surprisingly, the value of $R_{\rm H}$ is positive throughout the entire temperature range and shows a weak temperature dependence.
This clearly indicates that hole-type carriers are dominant for this compound, in spite of electron doping.
Moreover, the $x$=0.4 and 0.42 specimens also have the positive $R_{\rm H}$ value, although the $\rho_{yx}$ data are scattered due to their too high resistivity.
In contrast, for $x$=0.6, the $\rho_{yx}$ data exhibit a steep negative gradient, resulting in the large negative $R_{\rm H}$ of $-8.3\!\times\!10^{-2}$ cm$^{3}$/C.
Considering that the Bi impurity phase was detected for $x\!\ge\!0.5$ [Fig. \ref{fig:XRD}(a)], it is reasonable to ascribe the observed negative $R_{\rm H}$ to just extrinsic Hall signals from the Bi metal impurities, because a Bi metal has been reported to show huge negative $R_{\rm H}$ at low temperatures [For instance, $R_{\rm H}\!\sim\!-20$ cm$^{3}$/C at 4.2 K for polycrystalline Bi\cite{Brochin2001PRB}].
Actually, the temperature dependence of $R_{\rm H}$ for $x$=0.6 shows strong enhancement toward low temperatures (not shown here), which well coincides with the behavior of pure Bi metal.
It would be thus impossible to determine the sign of the intrinsic Hall coefficient for $x\!\ge\!0.5$, due to the likely contribution from the Bi impurities.\cite{Hallx05}
Note here that another impurity phase Bi$_{2}$S$_{3}$ has negative $R_{\rm H}$ as well\cite{Chen1997CM}, and hence the positive Hall signal should arise from the main phase of Sr$_{1-x}$La$_{x}$FBiS$_{2}$.
%
\par 
%
The existence of the hole-like Fermi surfaces in the normal state is consistent with the theoretical predictions.\cite{Usui2012PRB}
The calculations based on a two-orbital model (Bi 6$p_{x}$ and 6$p_{y}$) have revealed the topology of the Fermi surface dramatically changes, when approximately 0.5 electrons per Bi site are doped, i.e., at around $x$=0.5 in the present notation.
For the low-doping regime, electron pockets locate around the $K$ [($\pm\pi$,0) or (0,$\pm\pi$)] point.
When $x$ is increased up to $\sim$0.5, however, these pockets connect with each other and evolve into two hole pockets around the $\Gamma$ (0,0) and $M$ ($\pi$,$\pi$) points.
Thus, the observed positive $R_{\rm H}$ is considered to originate from the hole pockets formed after the topological transition of the Fermi surfaces.
On the other hand, the thermopower for $x$=0.45 exhibits a negative value, for instance, $\sim$-15 $\mu$V/K at room temperature (not shown here), which signals the coexistence of electron-type carriers with hole-type ones.
The observed small positive $R_{\rm H}$ may result from the cancellation of the contributions from the carriers with different signs, rather than large hole-like Fermi surfaces.
In theory, it was actually pointed out that new electron pockets show up again around the $K$ point with further increasing the carrier concentration above 0.5.\cite{Usui2012PRB}
Thus, our observations should put a strong constraint to the Fermi surface topology and relevant pairing mechanism for the BiS$_{2}$-based superconductor.
%
%
\section{Conclusions}
%
We have investigated the detailed electronic phase diagram and the superconducting properties for the electron-doped Sr$_{1-x}$La$_{x}$FBiS$_{2}$ layered compounds by synthesizing the polycrystalline samples with La concentration ($x$) systematically changed from 0 to 0.6.
Characterization using the powder x-ray diffraction and x-ray absorption spectroscopy has revealed that La substitution is successful up to $x$=0.6 and it effectively affords an electron into the BiS$_{2}$ layer.
The pristine compound is a band insulator and the resistivity systematically decreases with increasing $x$.
For $x\!\ge\!0.45$, the system exhibits metallic behavior except for the Anderson localization effect, where the superconducting transition manifests itself at low temperatures.
The bulk superconductivity was confirmed by the large diamagnetic response and clear jump in specific heat.
The variation in $T_{\rm c}$ is distinctive; it gradually increases with decreasing the carrier concentration toward the insulating phase and suddenly disappears at the critical concentration ($x\!\sim\!0.4$).
The obtained electronic phase diagram is characterized by the doping-induced insulator-superconductor transition with a steep phase boundary around $x$=0.4.
This markedly contrasts to the broad superconducting dome reported for the $R$O$_{1-x}$F$_{x}$BiS$_{2}$ compounds, which indicates the blocking layer strongly affects the superconducting transition.
Considering that all the bands arising from the Sr and F orbitals are away from the Fermi energy, the present phase diagram would purely reflect the effects of rigid-band-like electron doping into the BiS$_{2}$ layer.
%
\par
%
The thermodynamic measurements have furthermore suggested that the superconducting coupling strength is enhanced, as the phase boundary to the insulator is approached.
The observed superconducting features resemble those for Li-intercalated ZrNCl superconductors, where the transition to the superconducting phase takes place by as small as 6\% doping.
On the other hand, the small positive Hall coefficient as well as small negative thermopower was observed for the normal state of the present compound, which indicates that the topology of the Fermi surface significantly changes from that in the low-carrier density regime.
Therefore, the Sr$_{1-x}$La$_{x}$FBiS$_{2}$ system can be regarded as a layered superconductor derived from a doped band insulator, characterized by hole-like Fermi surfaces coexistent with electron-type ones.
%
\par
%
{\it Note added in proof}: We have recently become aware of a paper by Y. Li {\it et al.}\cite{Li2013arxive}, reporting on the phase diagram for Sr$_{1-x}$La$_{x}$BiS$_{2}$ ($0\!\le\!x\!\le\!0.7$).
%
\begin{acknowledgments}
%
This study was partly supported by Nippon Sheet Glass Foundation for Materials Science and Engineering, the Murata Science Foundation, the FIRST program by the Japan Society for the Promotion of Science, and JSPS Grant-in-Aid for Scientific Research (No. 23685014, No. 25620040, and No. 24224009).
The measurement in SPring-8 was performed under Proposal Number 2013A1902. 
%
\end{acknowledgments}
%


\begin{thebibliography}{99}
\bibitem{gamma} The values of $\gamma_{\rm n}$ for $x$=0.45, 0.5, and 0.6 are estimated from the specific heat data at 9 T to be 1.6, 1.3, and 1.4 mJ/mol K$^{2}$, respectively. The apparent volume fraction was extracted from the data in Fig. \ref{fig:RTMT}(a). The $\gamma_{\rm s}$ value for $x$=0.6 is approximately 0.9, which is reduced by approximately 30\% compared to that for $x$=0.5. This reduction roughly corresponds in size to that in the slope of $\Delta C/T$ ($\sim$20\%), which implies the coupling strength doesn't change significantly above $x$=0.5.
\bibitem{Hallx05} The Hall resistivity for $x$=0.5 is strongly curved as a function of $H$. This may be because the intrinsic positive signal from the main phase would be comparable to the negative signal from the Bi impurities.
\end{thebibliography}
\end{document}